\documentclass[12pt,oneside]{article}

\usepackage[margin=2cm]{geometry}

\usepackage{cite}
\usepackage{amsmath}
\usepackage{amsthm}
\usepackage{amssymb}
\usepackage{subfigure}
\usepackage{graphicx}
\usepackage{color}

\newtheorem{lemma}{Lemma}
\newtheorem{theorem}{Theorem}

\newcommand{\partsort}{{\sc Partial Order Production}}
\newcommand{\psort}{{\sc Partial Sorting}}

\newcommand{\sortwpi}{{\sc Sorting with Partial Information}}
\newcommand{\STAB}{\mathrm{STAB}}

\author{Jean Cardinal, Samuel Fiorini, Gwena\"el Joret\thanks{Universit\'e Libre de Bruxelles (ULB), Brussels, Belgium. {E-mail: \tt\small \{jcardin,sfiorini,gjoret\}@ulb.ac.be}
},\\ Rapha\"el M. Jungers\thanks{Universit\'e catholique de Louvain (UCL), Louvain-la-Neuve, Belgium. {E-mail: \tt\small raphael.jungers@uclouvain.be}  }, J. Ian Munro%
\thanks{University of Waterloo, Waterloo, Ontario, Canada. {E-mail: \tt\small imunro@uwaterloo.ca}}
}

\title{An Efficient Algorithm for Partial Order Production\footnote{This work was supported by the ``Actions de Recherche Concert\'ees'' (ARC) fund of the ``Communaut\'e fran\c{c}aise de Belgique'', NSERC of Canada, and the Canada Research Chairs Programme. G.J.\ and R.J.\ are Postdoctoral Researchers of the ``Fonds National de la Recherche Scientifique'' (F.R.S.--FNRS). A preliminary version of the work appeared in \cite{CFJJM09-stoc}.}}
\date{}
\begin{document}

\maketitle

\begin{abstract} 
We consider the problem of {\em partial order production}: arrange the elements of an unknown totally ordered set $T$ into a target partially ordered set $S$, by comparing a minimum number of pairs in $T$. Special cases include sorting by comparisons, selection, multiple selection, and heap construction.

We give an algorithm performing $ITLB + o(ITLB) + O(n)$ comparisons in the worst case. Here, $n$ denotes the size of the ground sets, and $ITLB$ denotes a natural information-theoretic lower bound on the number of comparisons needed to produce the target partial order.

Our approach is to replace the target partial order by a weak order (that is, a partial order with a layered structure) extending it, without increasing the information theoretic lower bound too much. We then solve the problem by applying an efficient multiple selection algorithm. The overall complexity of our algorithm is polynomial. This answers a question of Yao (SIAM J.\ Comput.\ 18, 1989).

We base our analysis on the entropy of the target partial order, a quantity that can be efficiently computed and provides a good estimate of the information-theoretic lower bound.
\end{abstract}

\section{Introduction}

We consider the \partsort\ problem:\smallskip

{\it Given a set $S = \{ s_1, s_2,\ldots , s_n\}$ partially ordered by a known partial order $\preccurlyeq$ and a set $T = \{ t_1, t_2,\ldots , t_n\}$ totally ordered by an unknown linear order $\leqslant$, find a permutation $\pi$ of $\{1,2,\ldots ,n\}$ such that $s_i \preccurlyeq s_j \Rightarrow t_{\pi (i)} \leqslant t_{\pi(j)}$, by asking questions of the form: ``is $t_i \leqslant t_j$?".}\smallskip

\begin{figure}
\begin{center}
\subfigure[Sorting]{\hspace{1cm}\includegraphics[scale=.15, angle=-90]{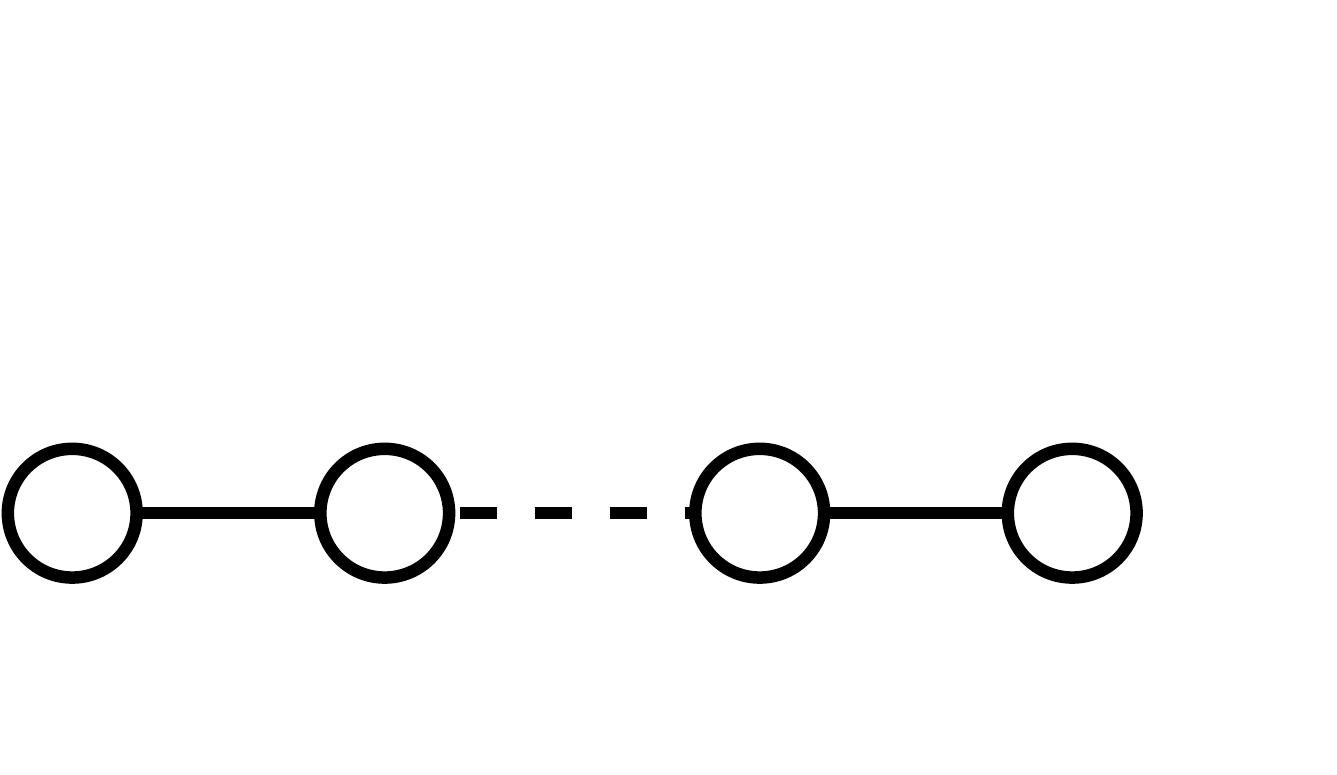}\hspace{1cm}}\hskip .5cm
\subfigure[Selection]{\hspace{1cm}\includegraphics[scale=.15, angle=-90]{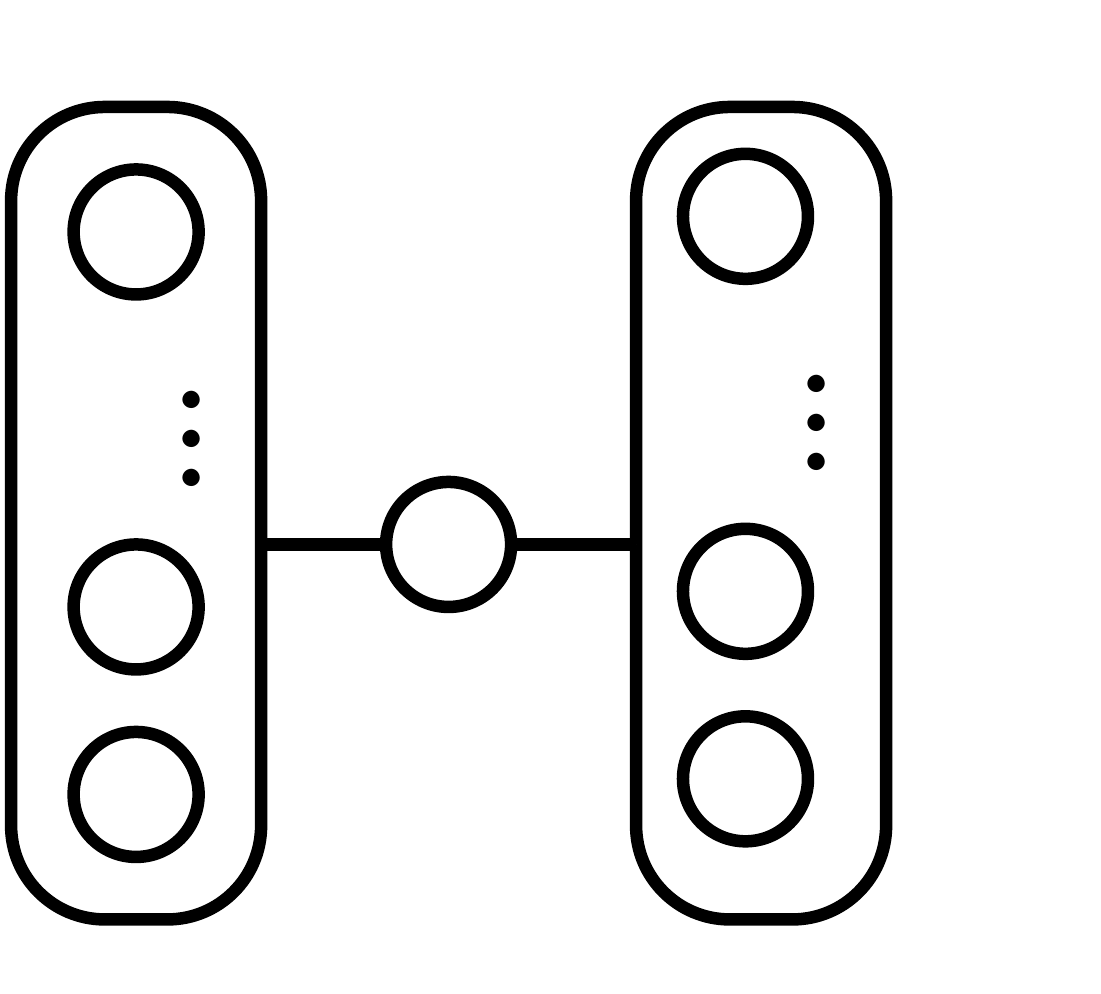}\hspace{1cm}}\hskip .5cm
\subfigure[Multiple selection]{\hspace{1cm}\includegraphics[scale=.15, angle=-90]{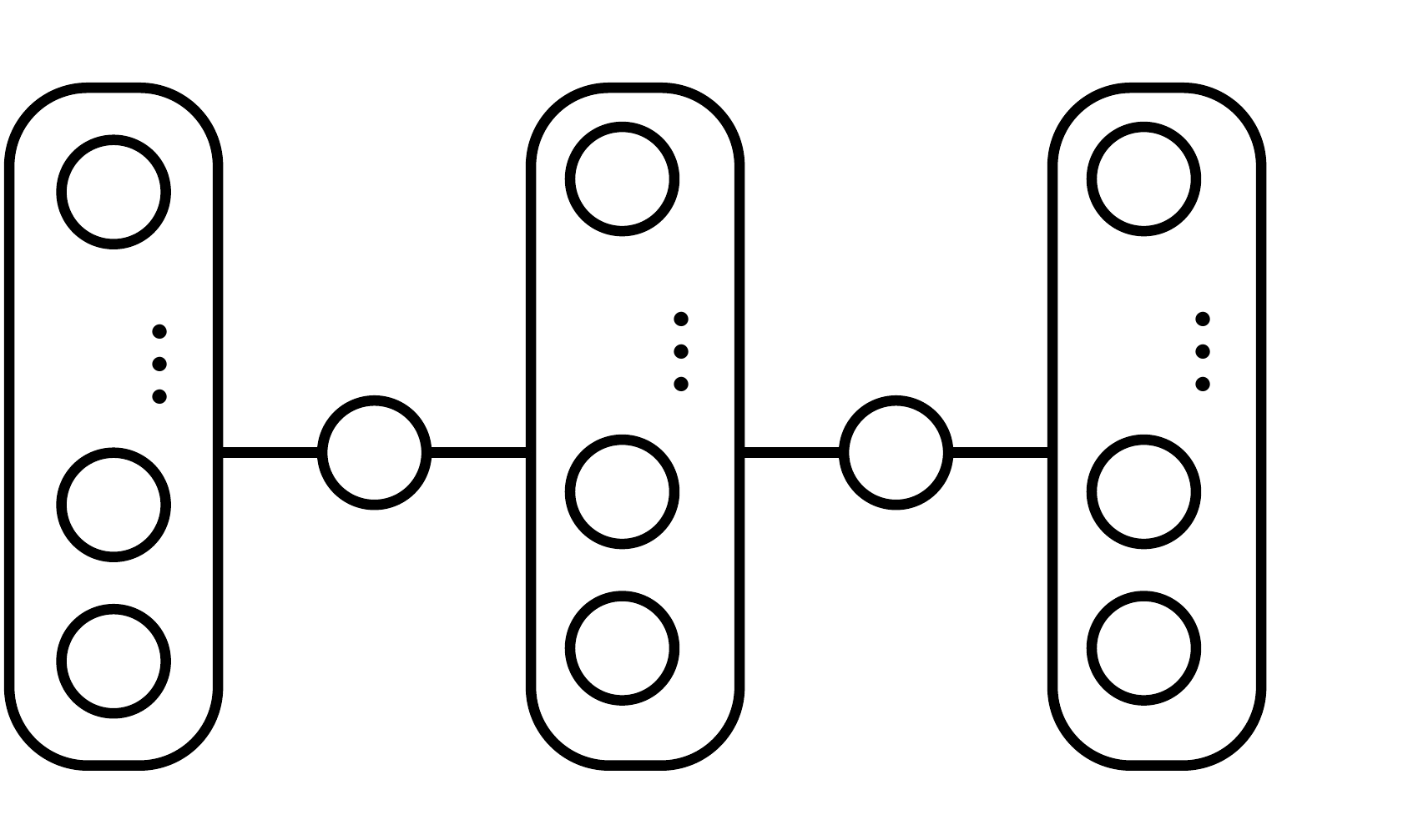}\hspace{1cm}}\hskip .5cm
\subfigure[Heap construction]{\hspace{1cm}\includegraphics[scale=.15, angle=-90]{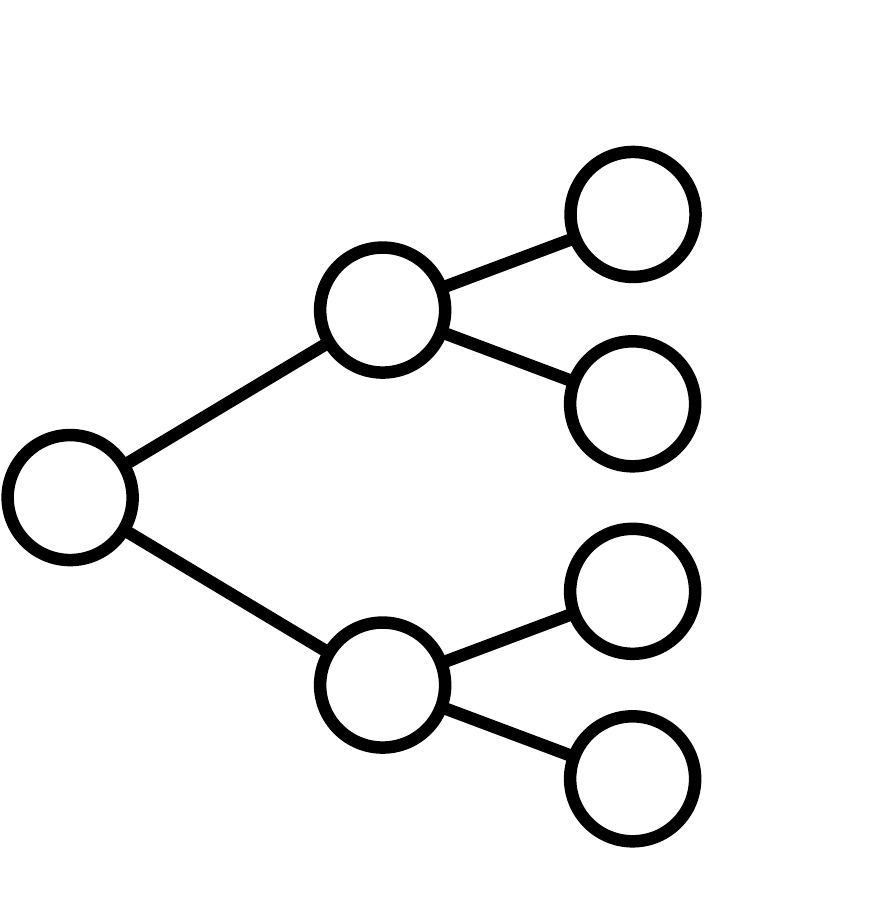}\hspace{1cm}}
\end{center}
\vspace{-.5cm}
\caption{\label{fig-special}Special cases of the \partsort\ problem.}
\end{figure}

The \partsort\ problem generalizes many fundamental problems (see Figure~\ref{fig-special}), corresponding to specific families of posets $P := (S,\preccurlyeq )$. It amounts to sorting by comparisons when $P$ is a chain. The selection~\cite{H61} and multiple selection~\cite{C71} problems are special cases in which $P$ is a weak order\footnote{Most of the terms that are not defined in the introduction are defined in Sections \ref{sec-perfect-graphs} and \ref{sec-algorithm}}, that is, has a layered structure (with $a \preccurlyeq b$ iff $a$ is on a lower layer than $b$). When the Hasse diagram of $P$ is a complete binary tree, the problem boils down to heap construction~\cite{CC92}.

We assume that the target poset $P$ is part of the input and represented by its Hasse diagram. Hence the size of the input can be $\Omega(n^2)$, whereas sorting the $n$ elements of $T$ takes $O(n \log n)$ time. In other words, reading the input could take more time than necessary to solve problem, provided a topological sorting of $P$ is known.

To cope with this paradoxical situation, we consider algorithms that proceed in two phases: a {\em preprocessing phase\/} during which an ordering strategy is determined (for instance, in the form of a decision tree, or any more efficient description, if possible), on the basis of the structure of $P$, and an {\em ordering phase\/} during which all comparisons between elements of $T$ are performed. Accordingly, we distinguish the {\em preprocessing complexity\/} and the {\em ordering complexity\/} of the algorithm, the latter being essentially proportional to the number of comparisons performed.

As noted before, we expect the overall complexity of the algorithm to be dominated by its preprocessing complexity. Thus it is desirable to perform the preprocessing phase only once, and then use the resulting ordering strategy on several data sets.


\paragraph{Lower bound on the number of comparisons}
We denote by $e(P)$ the number of linear extensions of the target poset $P$. Feasible permutations $\pi$ are in one-to-one correspondence with the linear extensions of $P$, thus the number of feasible permutations is exactly $e(P)$. On the other hand, the total number of permutations is $n!$. We have thus the following information-theoretic lower bound (logarithms are base 2):

\begin{theorem}[\hspace{-.01em}\cite{S76, A81, Y89}]
Any algorithm solving the \partsort\ problem for an $n$-element poset P requires
\begin{equation*}
ITLB := \log n! - \log e(P)
\end{equation*}
comparisons between elements of $T$ in the worst case and on average.
\end{theorem}

Note that we can assume without loss of generality that $P$ is connected, hence we also have a lower bound of $n-1$.

\paragraph{Problem history and contribution}
The \partsort\ problem was first proposed in 1976 by Sch\"onhage~\cite{S76}. It was studied five years after by Aigner~\cite{A81}. Another four years passed and the problem simultaneously appeared in two survey papers: one by Saks~\cite{S85} and the other by Bollob\'as and Hell~\cite{BH85}. In his survey, Saks conjectured that the \partsort\ problem can be solved by performing $O(ITLB) + O(n)$ comparisons in the worst case.

Four years later, in 1989, Yao proved Saks' conjecture~\cite{Y89}. He gave an algorithm solving the \partsort\ problem in at most $c_1\,ITLB + c_2\,n$ comparisons, for some constants $c_1$ and $c_2$. However, the preprocessing phase of Yao's algorithm seems difficult to implement efficiently. In fact, in the last section of his paper~\cite{Y89}, Yao asked whether, assuming $P$ is part of the input (as is the case here), there exists a polynomial-time algorithm for the problem that performs $O(ITLB) + O(n)$ comparisons.

Our main contribution is an algorithm that solves the \partsort\ problem and performs at most $ITLB + o(ITLB) + O(n)$ comparisons in the worst case. The preprocessing complexity of our algorithm is $O(n^3)$. Hence we answer affirmatively the question of Yao \cite{Y89} mentioned above. Moreover, we also significantly improve the ordering complexity, since Yao's constants $c_1$ and $c_2$ are quite large.

Further references, focussing mainly on lower bounds for the problem and generalizations of it include Culberson and Rawlins \cite{CR88}, Chen \cite{C94} and Carlsson and Chen \cite{CC94}.

\paragraph{Main ideas underlying our approach} We reduce the \partsort\ problem to the multiple selection problem. Instead of solving the problem for the given target poset $P$ we solve it for a larger (more constrained) poset that has a simpler structure, namely, a weak order $W$ extending $P$ (a weak order is a set of antichains with a total ordering between these antichains). This approach works because, as we show below, it is possible to find such a weak order $W$ whose corresponding information-theoretic lower bound $ITLB$ is not too large compared to that of $P$.

Unfortunately, computing $ITLB$ exactly is $\#P$-hard, because computing the number of linear extensions of a poset is $\#P$-complete, a result due to Brightwell and Winkler \cite{BW91}. The analysis is made possible because there exists a quantity, depending on the structure of the target poset, that can be computed in polynomial time and provides a good estimate of $ITLB$. This quantity is $nH$, where $H$ denotes the entropy of the considered target poset. (The entropy of a graph is defined in the next section, and the entropy of a poset is defined as the entropy of its comparability graph.) It was K\"orner who introduced the notion of the entropy of a graph, in the context of source coding~\cite{K73}. The idea of estimating an information-theoretic lower bound by means of the entropy of a poset was used before by Kahn and Kim in their inspiring work on sorting with partial information \cite{KK95}, see below.

\paragraph{Related problems}
In 1971 Chambers~\cite{C71} proposed an algorithm for the \psort\ problem, defined as follows: given a vector $V$ of $n$ numbers and a set $I\subseteq \{1,2,\ldots ,n\}$ of indices, rearrange the elements of $V$ so that for every $i\in I$, all elements with indices $j<i$ are smaller or equal to $V_i$, and elements with indices $j>i$ are bigger or equal to $V_i$. For the indices $i\in I$, the elements $V_i$ in the rearranged vector have rank exactly $i$, hence this problem is also called {\em multiple selection}. The \psort\ problem is a special case of \partsort\ in which the partial order is also a weak order.

The algorithm proposed by Chambers is similar to Hoare's ``find" algorithm~\cite{H61}, or QuickSelect. It has been refined and analyzed by Dobkin and Munro~\cite{DM81}, Panholzer~\cite{Pa03}, Prodinger~\cite{Pr03}, and Kaligosi, Mehlhorn, Munro, and Sanders~\cite{KMMS05}. For our purposes, the key result is that of Kaligosi {\em et al.\/}~\cite{KMMS05} in which it is shown that multiple selection can be done within a lower order term of the information theoretic lower bound, plus a linear term.\medskip

Another generalization of the sorting problem, called \sortwpi, was studied by Kahn and Kim~\cite{KK95}:\smallskip

{\it Given an unknown linear order $\leqslant$ on a set $T=\{ t_1,\ldots , t_n\}$, together with a subset $\preccurlyeq$ of the relations $t_i \leqslant t_j$ forming a partial order, determine the complete linear order $\leqslant$ by asking questions of the form: ``is $t_i \leqslant t_j$?".}\smallskip

This problem is equivalent to sorting by comparisons if $\preccurlyeq$ is empty. The information-theoretic lower bound for that problem is $\log e(Q)$, where $Q := (T,\preccurlyeq)$. The problem is complementary to the \partsort\ problem in the sense that sorting by comparisons can be achieved by first solving a \partsort\ problem, then solving the \sortwpi\ problem on the output.

A proof that there exists a decision tree achieving the lower bound up to a constant factor has been known for some time (see in particular Kahn and Saks~\cite{KS84j}). This is related to the $1/3$--$2/3$ conjecture of Fredman~\cite{F76} and Linial~\cite{L84}. Kahn and Kim~\cite{KK95} provided a polynomial time algorithm that finds the actual comparisons. They show that choosing the comparison that causes the entropy of $Q$ to increase the most leads to a decision tree that is near-optimal in the above sense.

\paragraph{Overview}
In Section~\ref{sec-perfect-graphs}, we study the entropy of perfect graphs. We show that it is possible to approximate the entropy of a perfect graph $G$ using a simple greedy coloring algorithm. More precisely, we prove that any such approximation is at most $H(G) + \log (H(G)+1) + O(1)$, where $H(G)$ denotes the entropy of graph $G$.\smallskip

Section~\ref{sec-algorithm} explains how to apply this result to solve the \partsort\ problem algorithmically. We begin the section by remarking that entropy is bound to play a central role for the problem since $nH(P) - n \log e \leq ITLB \leq nH(P)$, where $H(P)$ denotes the entropy of poset $P$.

The preprocessing phase of our algorithm starts by applying the greedy coloring algorithm studied in Section~\ref{sec-perfect-graphs} to the comparability graph of $P$.  We then modify this coloring (we ``uncross" the colors) in order to obtain an extension of $P$ which is an interval order $I$. Another application of the greedy coloring algorithm, this time on the comparability graph of $I$, yields a weak order $W$ extending $I$. Using our result on perfect graphs, we prove that the entropy of $W$ is not much larger than that of $P$, that is, $H(W) \le H(P) + 2\log (H(P)+1) + O(1)$.

The ordering phase of the algorithm simply runs then a multiple selection algorithm based on the weak order $W$. We use a multiple selection algorithm from Kaligosi {\it et al.}~\cite{KMMS05} that performs a number of comparisons close to the information-theoretic lower bound.

We conclude the section by proving that the preprocessing complexity of our algorithm is $O(n^3)$.\smallskip

Finally, in Section~\ref{sec-discussion}, we discuss {the number of comparisons and study} the existence of an algorithm solving the \partsort\ problem in $ITLB + O(n)$ comparisons. We give an example showing that such an algorithm cannot always reduce the problem to the case where the target poset is a weak order. More specifically, we exhibit a family of interval orders with entropy at most $\frac{1}{2}\log n$, any weak order extension of which has entropy at least $\frac{1}{2}\log n + \Omega(\log \log n)$.

\section{Entropy of Perfect Graphs}
\label{sec-perfect-graphs}
We recall that a subset $S$ of vertices of a graph is a \emph{stable set} (or {\em independent set}) if the vertices in $S$ are pairwise nonadjacent. Also, a graph $G$ is {\em perfect} if $\omega(H)=\chi(H)$ holds for every induced
subgraph $H$ of $G$, where $\omega(H)$ and $\chi(H)$ denote the
clique and chromatic numbers of $H$, respectively.

Let us recall similarly that the {\em stable set polytope\/} of an arbitrary graph $G$ with vertex set $V$ and order $n$ is the $n$-dimensional polytope
\begin{equation*}
\STAB(G) := \mathrm{\ conv} \{\chi^S \in \mathbb{R}^V : S \textrm{ stable set in }G\},
\end{equation*}
where $\chi^S$ is the characteristic vector of the subset $S$, assigning the value $1$ to every vertex in $S$, and $0$ to the others. The {\em entropy\/} of $G$ is defined as (see~\cite{K73,CKLMS90})
\begin{equation}
\label{def-H}
H(G) := \min_{x\in \STAB(G)} - \frac 1n \sum_{v\in V} \log x_v.
\end{equation}
For example, if $G=(V,E)$ is the graph with $V := \{a,b,c\}$ and $E := \{bc\}$, then $H(G) = 2/3$ and the minimum in (\ref{def-H}) is attained for $x = (x_a,x_b,x_c) = (1,1/2,1/2)$.

Note that graph entropy was originally defined with respect to a given probability distribution on $V$. However, for our purposes we can take the uniform distribution, as in \cite{KK95}. In this case we obtain Equation (\ref{def-H}).

An upper bound on $H(G)$ can be found as follows: First, use the greedy coloring algorithm that removes iteratively a maximum stable set from $G$, giving a sequence $S_{1}, S_{2}, \dots, S_{k}$ of stable sets of $G$. If $G$ is perfect, this can be done in polynomial time (see, e.g., Gr\"otschel, Lov\'asz and Schrijver \cite{GLS93}). Next, let $\tilde x \in \mathbb{R}^V$ be defined as
$$
\tilde x := \sum_{i=1}^k \frac{|S_{i}|}{n} \cdot \chi^{S_{i}}.
$$
By definition, $\tilde x \in \STAB(G)$. We call any such point $\tilde{x}$ a {\em greedy point}. The value of the objective function in the definition of $H(G)$ for $\tilde x$ is $\sum_{i=1}^k \frac{|S_{i}|}{n} \log \frac{n}{|S_{i}|}$. We refer to the latter quantity simply as the entropy of $\tilde x$.
It turns out that this gives a good approximation of $H(G)$ when $G$ is a perfect graph.

\begin{theorem}
\label{th-greedy-perfect-graphs}
Let $G$ be a perfect graph on $n$ vertices and denote
by $\tilde{g}$ the entropy of an arbitrary greedy point $\tilde{x} \in \STAB(G)$. Then
$$
\tilde{g} \leq \frac{1}{1-\delta} \left(H(G) + \log \frac{1}{\delta} \right)
$$
for all $\delta > 0$, and in particular
$$
\tilde{g} \leq H(G) + \log (H(G)+1) + O(1).
$$
\end{theorem}

A key tool in our proof of Theorem~\ref{th-greedy-perfect-graphs} is a min-max relation of Csisz{\'a}r, K{\"o}rner, Lov{\'a}sz, Marton, and Simonyi~\cite{CKLMS90} relating the entropy of a perfect graph $G$ to the entropy of its complement $\bar G$:

\begin{theorem}[\hspace{-.01em}\cite{CKLMS90}]
\label{th-sum-log-n}
If $G$ is a perfect graph on $n$ vertices, then $H(G) + H(\bar G)=\log n$.
\end{theorem}

We now turn to the proof of Theorem~\ref{th-greedy-perfect-graphs}.

\begin{proof}[Proof of Theorem~\ref{th-greedy-perfect-graphs}]
Let $S_{1}, S_{2}, \dots, S_{k}$ be the sequence of stable sets of $G$ selected by the greedy
algorithm (in the order the algorithm removes them).
So $S_1$ is a maximum stable set in $G$, $S_2$ is a maximum stable set in $G - S_1$, and so on.
The outline of the proof is as follows: We first use the sets
$S_{1}, S_{2}, \dots, S_{k}$  to define a point $z \in \mathbb{R}^{V}$, where $V$ is the vertex set of $G$. We then
show that  $z$ belongs to the stable set polytope of the {\em complement} $\bar G$ of $G$, that
is, $z \in \STAB(\bar G)$. Finally, we derive the desired inequality by combining the
upper bound on $H(\bar G)$ implied by $z$ with Theorem~\ref{th-sum-log-n}.

Fix $\delta>0$. For each vertex $v \in V$ we let $m = m(v)$ be the unique index in $\{1,\ldots,k\}$
such that $v \in S_m$. We define $z$ by letting, for each vertex $v$ of $G$,
$$
z_{v} := \frac{\delta}{n} \left(\frac{1}{\tilde{x}_v}\right)^{1-\delta}
= \frac{\delta}{n} \left(\frac{n}{|S_{m(v)}|}\right)^{1-\delta}
= \frac{\delta}{n^\delta} \left(\frac{1}{|S_{m(v)}|}\right)^{1-\delta}.
$$
We claim that for every stable set $S$ of $G$:
\begin{equation}
\label{eq-clique-n}
\sum_{v \in S} z_{v} \leq 1.
\end{equation}

Write the stable set $S$ as $S = T_{1} \cup T_{2} \cup \cdots \cup T_{\ell}$, where $T_{i}$ is the $i$th subset of $S$ taken by the greedy algorithm during its execution.
For every $v \in T_{1}$, we have  $S_{m(v)}=S_1,$ and $|S_{m(v)}| \geq |S|$, since the greedy algorithm could have
selected the set $S$ when it took $S_{m(v)}$.
More generally, for every $i \in \{1,2, \dots, \ell\}$ and $v \in T_{i}$, we have
$|S_{m(v)}| \geq |S| - \sum_{j=1}^{i-1}|T_{j}|$. It follows in particular that we can enumerate the points of $S$ as $v_1$, $v_2$, \ldots, $v_s$ in such a way that
$$
|S_{m(v_{i})}| \geq |S| - i +1 \qquad \forall i \in \{1,2,\ldots,s\}.
$$
We thus have
\begin{eqnarray*}
\sum_{v \in S} z_{v}
&\le &
\frac{\delta}{n^\delta} \left(\left(\frac{1}{|S|}\right)^{1-\delta} + \left(\frac{1}{|S|-1}\right)^{1-\delta} + \ldots + 1\right)\\
&\le &
\frac{\delta}{n^\delta} \left(\int_0^{|S|} \frac{1}{x^{1-\delta}} \mathrm{d}x\right)\\
&\le & 1.
\end{eqnarray*}
Equation~\eqref{eq-clique-n} follows.

Two classical results on perfect graphs are that the stable set polytope is completely described
by the non-negativity and clique inequalities, that is,
$$
\STAB(G) = \{x \in \mathbb{R}^V_+ : \sum_{v \in K} x_v \le 1\quad \forall K \textrm{ clique in } G\}
$$
(see Chv\'atal~\cite{C75}), and that the complement $\bar G$ of $G$ is also a
perfect graph (Lov\'asz~\cite{L72}). Combining these two results with~\eqref{eq-clique-n}
shows that $z \in \STAB(\bar G)$.
Using Theorem~\ref{th-sum-log-n}, we then deduce
\begin{eqnarray*}
H(G) &= & \log n - H(\bar G)\\
&\ge & \log n + \frac{1}{n}\sum_{v\in V}\log z_{v}\\
&= & \log n + \frac{1}{n}\sum_{v\in V}\log\left( \frac{\delta}{n} \left(\frac{1}{\tilde{x}_v}\right)^{1-\delta}\right)\\
&= &- \frac{1-\delta}{n} \sum_{v \in V} \log \tilde{x}_v - \log \frac{1}{\delta}\\
&= & (1-\delta) \tilde{g} - \log \frac{1}{\delta}.
\end{eqnarray*}
Hence, $\tilde{g} \leq \frac{1}{1-\delta} \left(H(G) + \log \frac{1}{\delta} \right)$, for all $\delta>0$.
By choosing $\delta = 1/2$ if $H(G)\leq 1$, and $\delta =1/(H(G) + 1)$ otherwise, we obtain
$\tilde{g} \leq H(G) + \log (H(G) + 1) + O(1)$.
\end{proof}

\section{An Algorithm for Partial Order Production}
\label{sec-algorithm}

We denote by $G(P)$ the comparability graph of a poset $P = (V,\leqslant_P)$, and let $H(P) := H(G(P))$. Note that a stable set in $G(P)$ is an antichain in $P$, that is, a set of mutually incomparable elements. Note also that $G(P)$ is perfect, a basic result that is dual to Dilworth's theorem, see, e.g., \cite{G04}. The relevance of the notion of graph entropy in the context of sorting was first observed by Kahn and Kim~\cite{KK95}. Using the fact that the volume of $\STAB(G(P))$ equals $e(P)/n!$ (see Stanley~\cite{S86}), they proved the following result.

\begin{lemma}[\hspace{-.01em}\cite{KK95}]
For any poset $P$ of order $n$,
\begin{equation*}
- nH(P) \le \log e(P) - \log n! \le n \log n - \log n! - n H(P).
\end{equation*}
\end{lemma}

When written as
$$
2^{-nH(P)} \le \frac{e(P)}{n!} \le 2^{-nH(P)} \cdot \frac{n^n}{n!},
$$
the above inequalities become intuitively clear, since $2^{-nH(P)}$ is the (maximum) volume of a box contained in $\STAB(G(P))$, $e(P) / n!$ is the volume of $\STAB(G(P))$, and $2^{-nH(P)} \cdot n^n/n!$ is the volume of a simplex containing $\STAB(G(P))$. The lemma directly implies the following equality for every poset $P$:
\begin{equation}
\label{eq-H}
ITLB=\log n! - \log e(P) = n H(P) + O(n).
\end{equation}

We recall that a poset is said to be a {\em weak order\/} whenever its comparability graph is a complete
$k$-partite graph, for some $k$. Such a poset $W = (V,\leqslant_W)$ can be partitioned into $k$ maximal
antichains $A_1$, \ldots, $A_k$, the {\em layers\/} of $W$, such that $v <_W w$ whenever there exist indices $i$
and $j$ such that $v \in A_i$, $w \in A_j$ and $i < j$. When restricted to weak orders, the \partsort\ problem
resembles the \psort\ problem, with $I = \{\sum_{j=1}^i |A_j| : i=1,\ldots, k-1\}$.

Our key idea is to show that, using (twice) the greedy coloring algorithm presented in the previous section, we can efficiently extend\footnote{A poset $Q$ {\em extends\,} a poset $P$ if
they have the same ground set $V$ and $v \leqslant_P w$ implies $v \leqslant_Q w$, for all $v, w \in V$.}
the given poset $P$ to a weak order $W$ whose entropy is close to that of $P$.  The reason why we have to use twice the greedy algorithm is that the obtained coloration might not be "ordered" (might not represent the stable sets of a weak order).  However, we describe below how to uncross this coloring in order to extend $P$ to an interval order without increasing too much the entropy.  We show that applying our greedy coloring to an interval order provides an "ordered" coloring, which allows us to run a second time our greedy algorithm, providing an extension which is a weak order.

We then simply run an efficient multiple selection procedure, with $W$ as input.
We show that, because replacing $P$ by $W$ does not increase the entropy too much, the resulting number of comparisons is close to $ITLB$.

\begin{figure}
\begin{center}
\subfigure[\label{fig-coloring} Possible greedy point $\tilde{x}$.]{\includegraphics[scale=.25, angle=-90]{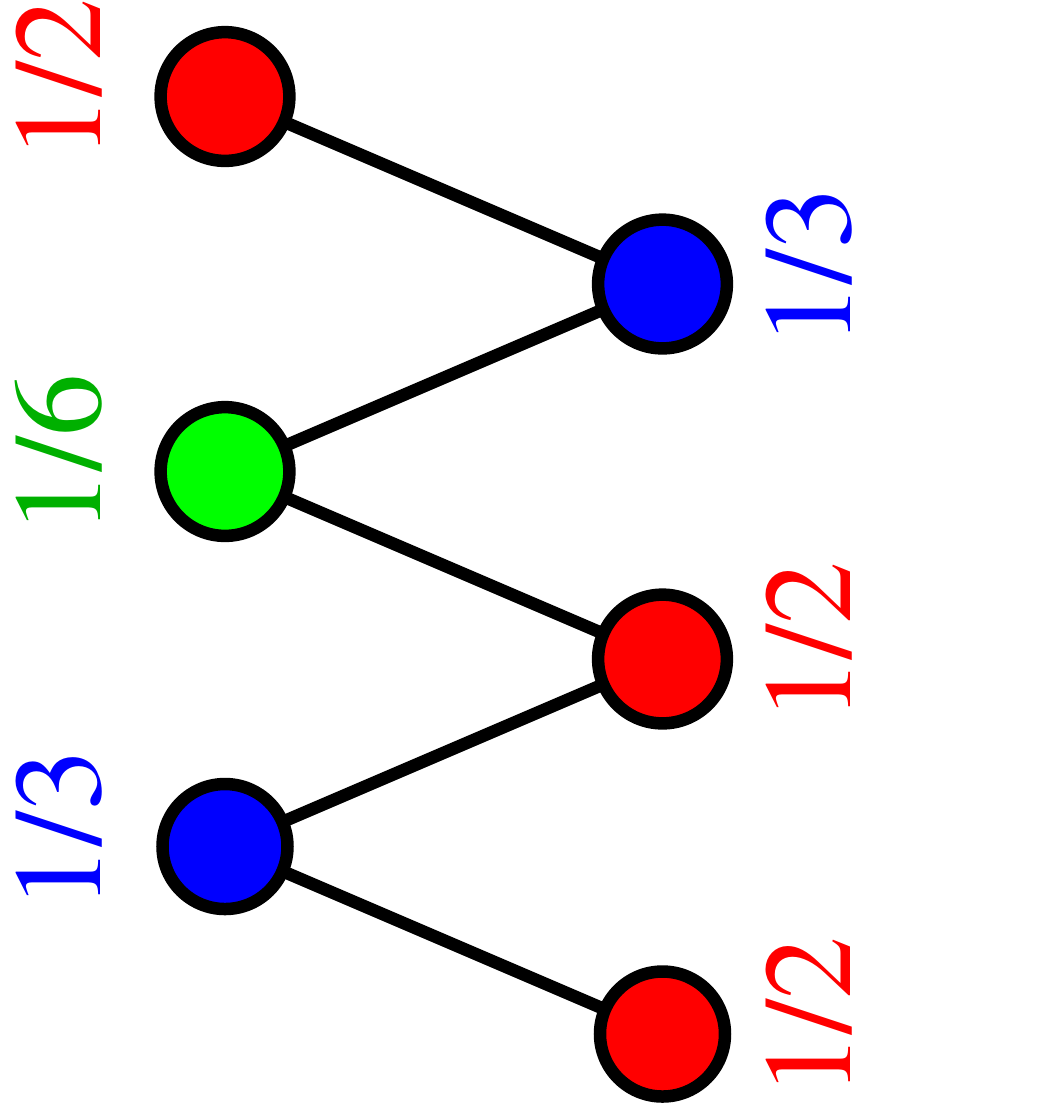}}\hskip 2.5cm
\subfigure[\label{fig-network} Network $D$ and potential $\tilde{y}$.]{\includegraphics[scale=.25, angle=-90]{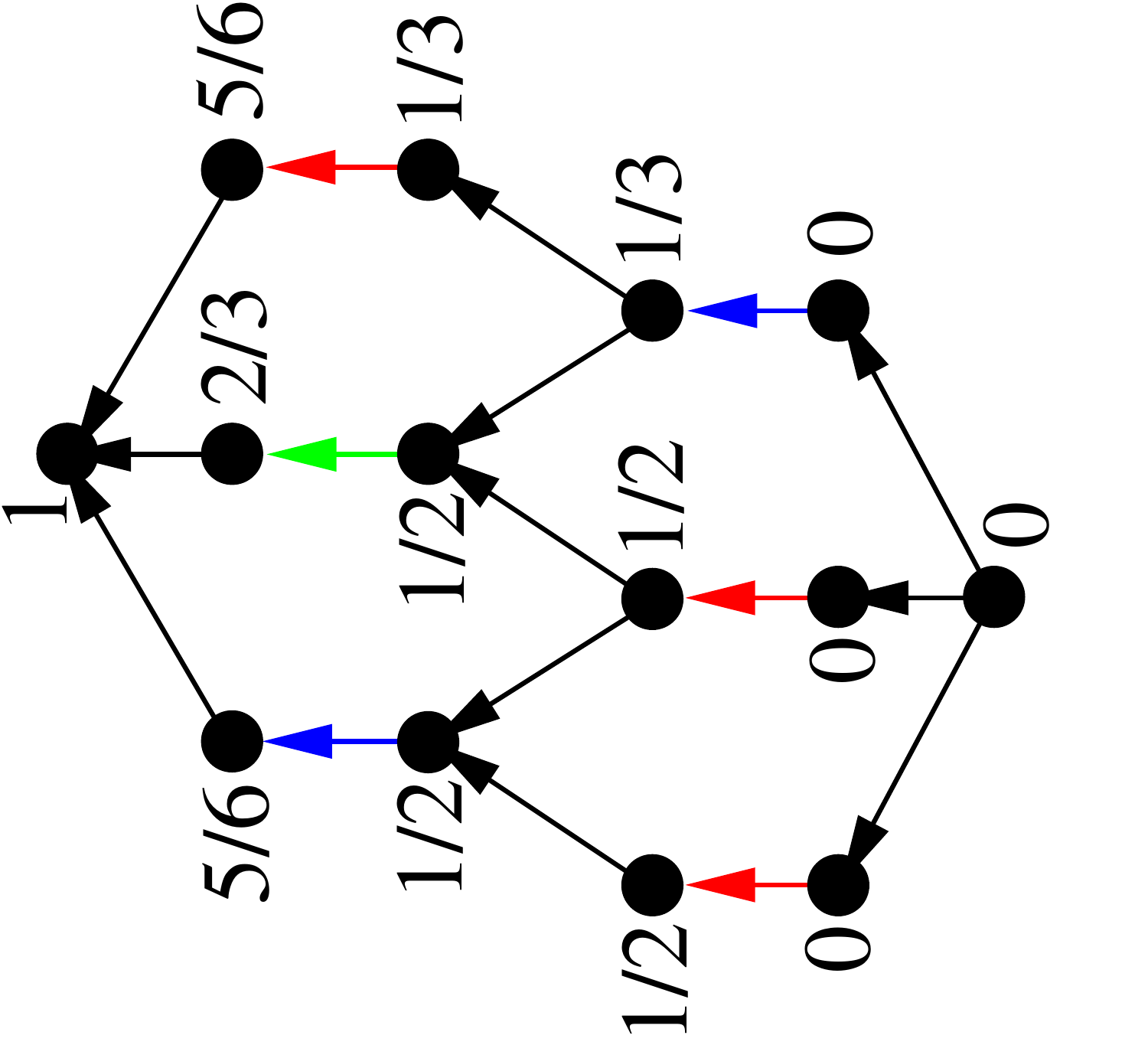}}\\
\subfigure[\label{fig-intervals} Interval representation of $I$.]{\includegraphics[scale=.25, angle=-90]{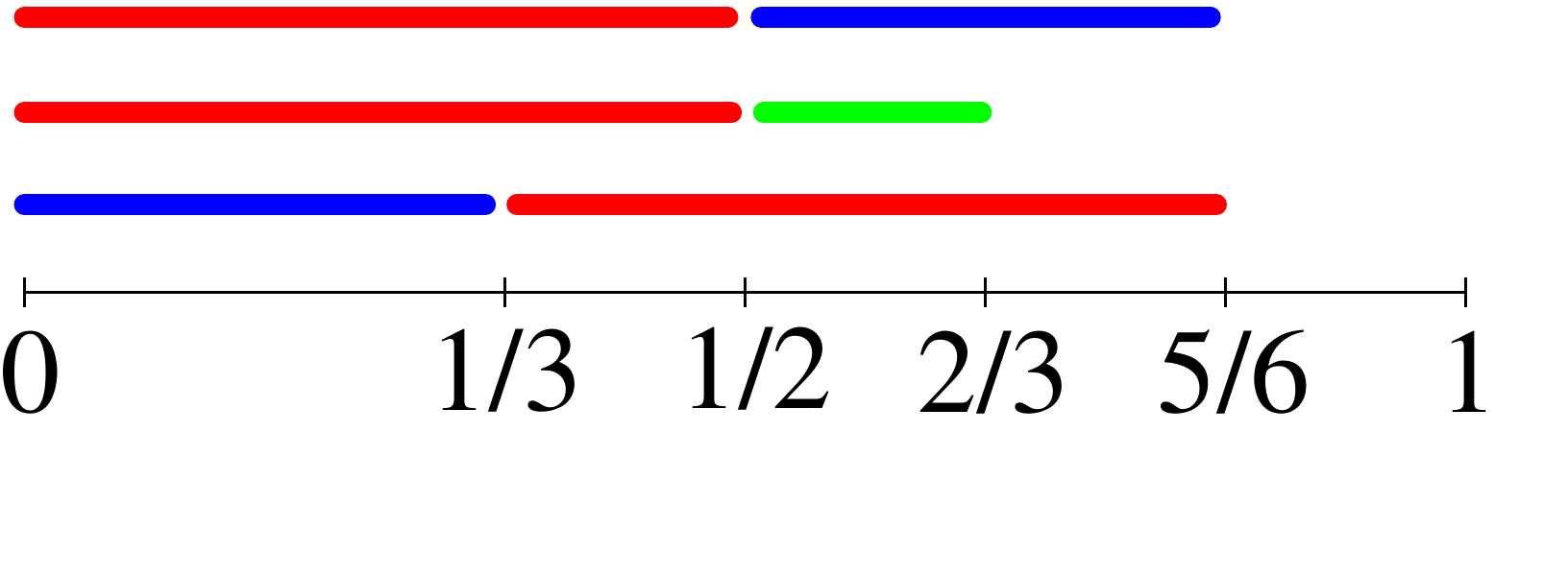}}\hskip 2cm
\subfigure[\label{fig-intervalgraph} Interval order $I$.]{\includegraphics[scale=.25, angle=-90]{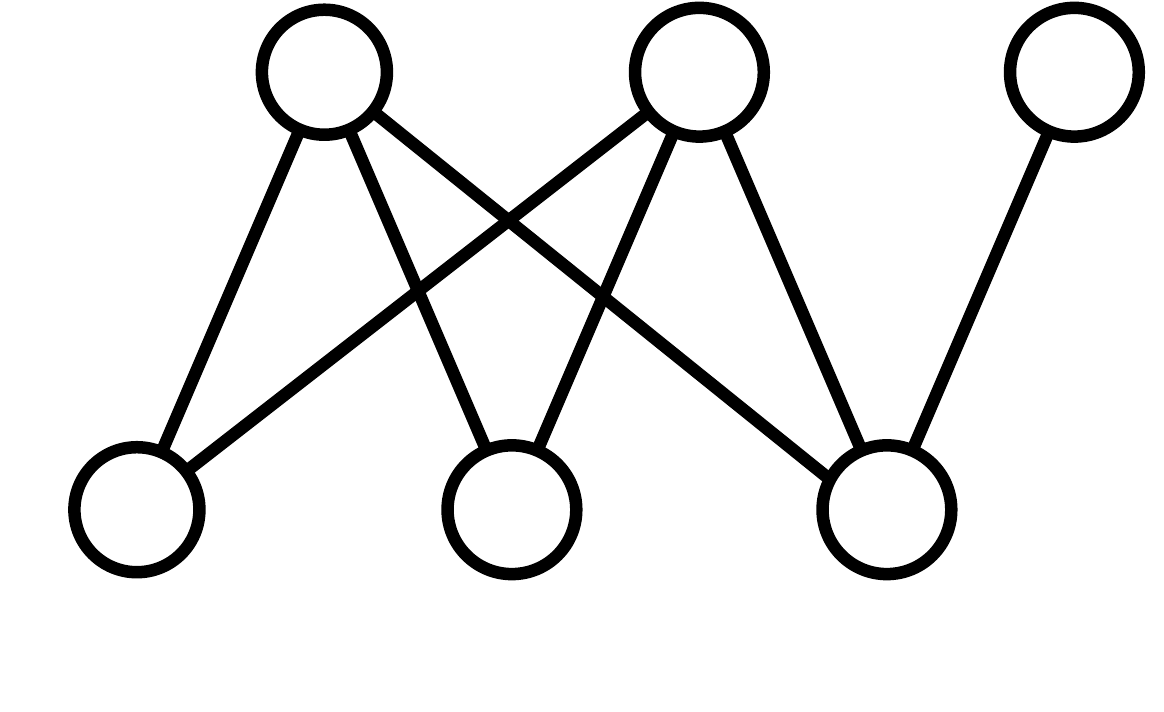}}
\end{center}
\caption{\label{fig-weakextension} Obtaining an interval order extension of the poset $P$.}
\end{figure}

The preprocessing phase is composed of three steps, each of which can be performed in polynomial time. In the first step, we apply the greedy coloring procedure to $G(P)$, to obtain a greedy point $\tilde x$. This step makes use of an auxiliary network defined from $P$. Then, in the second step, using again the auxiliary network, we extend $P$ to an interval order $I$ whose entropy is not larger than that of $\tilde x$. This allows us to ``uncross" the antichains used in $\tilde x$. (An alternative way of obtaining the interval order $I$ is to apply Kahn and Kim's \cite{KK95} laminar decomposition lemma to $\tilde{x}$.) Finally, in the third step, we apply the greedy coloring procedure again, this time on $G(I)$, to obtain the weak order $W$. See Figure \ref{fig-weakextension} for an illustration of steps 1 and 2.

\paragraph{Auxiliary network}
Let $P = (V,\leqslant_P)$ be any poset. We say that $v$ is {\em covered\/} by $w$ in $P$ if $v \leqslant_P w$, $v \neq w$ and $v \leqslant_P z \leqslant_P w$ implies $z = v$ or $z = w$. The {\em Hasse diagram\/} of $P$ is the network with node set $V$, and arc set $\{(v,w) : v$ is covered by $w$ in $P\}$. An element $v$ of $P$ is {\em minimal\/} (resp.\ {\em maximal\/}) if $z \leqslant_P v$ (resp.\ $v \leqslant_P z$) implies $z = v$.

We construct a network $D = D(P)$ from the Hasse diagram of $P$ by first uncontracting each element $v \in V$ to an arc $(v^-,v^+)$ and then adjoining a source node $s$ sending an arc to each minimal element, and a sink node $t$ receiving an arc from each maximal element. The resulting network has node set
$$
N(D) := \{s,t\} \cup \{v^- : v \in V\} \cup \{v^+ : v \in V\}
$$
and arc set
$$
\begin{array}{r@{\,\,}c@{\,\,}l}
A(D) &:= &\{(s,v^-) : v \in V,\ v \textrm{ minimal in } P\} \cup \{(v^-,v^+) : v \in V\} \cup \mbox{}\\[.5ex]
&&\{(v^+,w^-) : v \textrm{ is covered by } w \textrm{ in } P\} \cup \{(v^+,t) : v \in V,\ v \textrm{ maximal in } P\}.
\end{array}
$$
This network gives a useful characterization of points in the stable set polytope of the comparability graph of $P$, as is explained in the next lemma.

\begin{lemma}
\label{lem-potential}
Let $P$ be a poset with ground set $V$, let $G:=G(P)$ and $D:=D(P)$. A vector $x \in \mathbb{R}^V$ belongs to $\STAB(G)$ if and only if there exists a vector $y \in \mathbb{R}^{{N(D)}}$ (called a \emph{potential\/}) such that $y_{s} = 0$, $y_{t} = 1$, $y$ is nondecreasing along arcs of $D$, and $y_{v^+} - y_{v^-} = x_v$ for all $v \in V$.
\end{lemma}
\begin{proof}
Again, we use (see Chv\'atal~\cite{C75}):
$$
\STAB(G) = \{x \in \mathbb{R}^V_+ : \sum_{v \in K} x_v \le 1\quad \forall K \textrm{ clique in } G\}.
$$

We first show sufficiency. Let $x \in \mathbb{R}^V$ be a vector that admits a potential $y \in \mathbb{R}^{N(D)}$. Consider any chain $C = \{v_1,v_2,\ldots,v_c\}$ in $P$ with $v_1 \leqslant_P v_2 \leqslant_P\cdots \leqslant_P v_c$ (cliques in $G$ correspond to chains in $P$). Then
\begin{eqnarray*}
\sum_{v \in C} x_v &= & (y_{v_1^+}-y_{v_1^-}) + \cdots + (y_{v_c^+}-y_{v_c^-})\\
&\le &(y_{v_1^-}-y_{s}) + (y_{v_1^+}-y_{v_1^-}) +
(y_{v_2^-} - y_{v_1^+}) + \cdots + (y_{v_c^+}-y_{v_c^-}) + (y_{t} - y_{v_c^+})\\[2ex]
&= &y_{t} - y_{s} = 1.
\end{eqnarray*}
It follows that $x \in \STAB(G)$.

For necessity, consider $x \in \STAB(G)$. For $v \in V$, we let $y_{v^+}$ be the maximum total weight of a chain of $P$ whose maximum with respect to $\leqslant_{P}$ is $v$,
when each vertex $w$ is given the weight $x_w$, and $y_{v^-} := y_{v^+} - x_v$. Then we let $y_{s} := 0$ and $y_{t} := 1$. As is easily verified, $y$ is a potential for $x$.
\end{proof}

It follows that $H(P)$ is the optimum value of the following convex minimization problem with a polynomial number of variables and constraints:
$$
\begin{array}{rrrcll}
\textrm{\rm (H-potential)}
&\min &\multicolumn{3}{l}{\displaystyle -\frac{1}{n} \sum_{v \in V} \log x_v}\\
&\textrm{\rm s.t.} &x_v  &= &y_{v^+} - y_{v^-} &\forall v \in V\\[.5ex]
&                  &y_{p}   &\leqslant &y_{q} &\forall (p,q) \in A(D)\\
&                  &y_{s} &= &0\\
&                  &y_{t}  &= &1.
\end{array}
$$
We remark that this formulation shows that $H(P)$ can be computed to within any fixed precision in strongly polynomial time, using interior point methods (see for instance~\cite{NN94}). However, approximating  $H(P)$ using a greedy point will be enough for our purposes, and will moreover give a better upper bound on the complexity of our algorithm.

\paragraph{Greedy extensions}
Let $\tilde x$ be a greedy point in $\STAB(G)$, as defined in Section~\ref{sec-perfect-graphs}. Consider the potential $\tilde y\in \mathbb{R}^{N(D)}$ defined from $\tilde x$ as in the proof of Lemma~\ref{lem-potential}: For $v\in V$, we let $\tilde y_{v^{+}}$ be the maximum (total) weight of a chain of $P$ ending in $v$, where each vertex $w$ has weight $\tilde x_{w}$, and $\tilde y_{v^{-}} := \tilde y_{v^{+}} - \tilde x_{v}$. Let also $\tilde y_{s} :=0$ and $\tilde y_{t} :=1$.

From this potential $\tilde y$, we compute an interval order $I$ extending $P$ whose entropy is not larger than that of $\tilde x$. The ground set of $I$ is $V$. We let $v \leqslant_I w$ whenever $\tilde{y}_{v^+} \le \tilde{y}_{w^-}$. Thus the open intervals $(y_{v^-},y_{v^+})$ (for $v \in V$) provide an interval representation of $I$. Because $v \leqslant_P w$ implies $\tilde{y}_{v^+} \le \tilde{y}_{w^-}$, which in turn implies $v \leqslant_I w$, the interval order $I$ extends $P$.  The entropy of $I$ is not larger than that of $\tilde x$ because $(\tilde x,\tilde y)$ remains feasible for the minimization problem (H-potential) defined above, after  $P$ is replaced by $I$.

Apply again the greedy coloring algorithm, but now on $G(I)$. Let $A_1$, \ldots, $A_k$ denote the antichains of $I$ produced by the greedy coloring algorithm. Because $I$ is an interval order, we can find a permutation $\sigma$ of $\{1,\ldots,k\}$ such that $v <_I w$, $v \in A_{\sigma(i)}$ and $w \in A_{\sigma(j)}$ imply $i < j$. Thus, the weak order $W$ with ground set $V$ obtained by setting $v <_W w$ whenever $v \in A_{\sigma(i)}$ and $w \in A_{\sigma(j)}$ with $i < j$ is an extension of $I$. Such a weak order $W$ is said to be a {\em greedy extension\,} of the original poset $P$.

\begin{lemma}
\label{lem-greedy}
Let $P$ be a poset and $W$ one of its greedy extensions.
Then
$$
H(W) \le \frac{1}{1-\delta} \left(H(P) + 2\log \frac{1}{\delta} + 2\right)
$$
for all $\delta > 0$, and in particular
$$
H(W) \le H(P) + 2\log (H(P)+1) + O(1).
$$
\end{lemma}
\begin{proof}
Let $\delta' := \delta/2$. Let $I$ denote the intermediate interval order used to obtain $W$. Theorem~\ref{th-greedy-perfect-graphs} implies
\begin{align*}
H(P) &\geq (1-\delta')H(I) - \log(1/\delta') \\
&\geq(1-\delta')\big( (1-\delta')H(W) - \log(1/\delta') \big) - \log(1/\delta') \\
&\geq (1-\delta) H(W) - 2 \log(1/\delta) - 2.
\end{align*}
In addition to Theorem~\ref{th-greedy-perfect-graphs}, for the first inequality we used the fact that $H(I)\leq \tilde g,$ and for the second one, we used the fact that the greedy coloring of $I$ directly gives the unique decomposition of $W$ in maximal stable sets.
This shows the first part of the claim. For the second part, again take $\delta = 1/2$ if $H(P)\leq 1$, and $\delta =1/(H(P)+1)$ otherwise.
\end{proof}

\paragraph{Algorithm {and complexity}}
The above results directly suggest the following algorithm:
compute a greedy extension $W$ of $P$, and
run a multiple selection procedure on $T$ with respect to $W$. In terms of the number of comparisons
between elements of $T$, we only incur a controlled penalty.

\begin{theorem}
\label{th-ub}
The \partsort\ problem can be solved in polynomial time using at most
\begin{equation}
\label{ub-exp}
ITLB + o(ITLB) + O(n)
\end{equation}
comparisons between elements of $T$ in the worst case.
\end{theorem}
\begin{proof}
The weak order extension $W$ can be computed in polynomial time.
Let us denote by $A_1$, \ldots, $A_k$ its layers. We run the multiple selection algorithm on the elements of $T$, with the ranks $r_i := \sum_{j=1}^i |A_j|$ (for $i=1, \ldots, k-1$). Kaligosi {\it et al.}~\cite{KMMS05} give a multiple selection algorithm that requires only
$B + o(B) + O(n)$ comparisons in the worst case, where $B := \log n! - \log e(W)$ is the information-theoretic lower
bound for $W$. Thus
\begin{eqnarray*}
B  & =    & n H(W) + O(n) \hspace{7.2em} {\textrm{ (from\ Eqn.~(\ref{eq-H}))}}\\
   & \leq & n H(P) + 2 n \log (H(P)+1) + O(n) \hspace{1em} \textrm{ (from Lemma~\ref{lem-greedy})} \\
   & = & ITLB + 2 n \log \left( \frac{ITLB}{n} + 1\right) + O(n)\hspace{.9em} {\textrm{ (from\ Eqn.~(\ref{eq-H}))}}  \\
   & = & ITLB + o(ITLB) + O(n).
\end{eqnarray*}
Hence $B + o(B) + O(n) = ITLB + o(ITLB) + O(n)$, and the theorem follows.
\end{proof}

We conclude the section by discussing the preprocessing complexity of our algorithm.

The first execution of the greedy coloring algorithm can be done in time $O(mn)$,
where $m$ is the number of arcs in the network $D:=D(P)$ (notice $m\geq n$ and $m = O(n^2)$), as we now briefly explain. The algorithm finds maximal antichains in the graph by decrementing a flow on the auxiliary network.  This flow has to satisfy lower bounds on the arcs.

Let $X := \varnothing$, $i:=1$, and put a lower bound
of $\ell_{a}:=1$ on each arc $a$ of the form $(v^{-}, v^{+})$ with
$v \in V $, of $\ell_{a}:=0$ on every other arc $a$ of $D$.
Start with an arbitrary integer $s$--$t$ flow $\phi$ of value $n$ such that $\phi_{a} \geq \ell_{a}$
for every arc $a\in A(D)$. 
Let $Y$ be the set of nodes of $D$ that can be reached
from $s$ following a {\em decrementing path}, namely, a
path $v_{0}v_{1}\dots v_{k}$ with $v_{0}:=s$ such that, for every $i \in \{1, 2, \dots, k\}$,
either $(v_{i-1}, v_{i}) \in A(D)$ and $\phi_{(v_{i-1}, v_{i})} > \ell_{(v_{i-1}, v_{i})}$, or
$(v_{i}, v_{i-1}) \in A(D)$. Now, there are two cases: (1)  $t \in Y$. Thus there exists
a decrementing $s$--$t$ path. We then decrement by $1$ the flow value of $\phi$
using the latter path. (2) $t \notin Y$. Observe that no arc of $D$ enters the set $Y$ and that
the arcs $a$ going out of $Y$ satisfy $\phi_{a} = \ell_{a}$.
It follows that
$$
A_{i} := \{v\in V \mid (v^{-}, v^{+}) \in \delta^{+}(Y),\ \phi_{(v^{-}, v^{+})} = 1\}
$$
is an antichain of $P-X$. (Here,  $\delta^{+}(Y)$ denotes the set of arcs of $D$ going out of $Y$.)
Moreover, since the flow value of $\phi$ equals $|A_{i}|$, the antichain $A_{i}$
is maximum among the antichains of $P - X$. This is because, by definition of our lower bounds,
the flow value is at least $|A|$, for every antichain $A$ contained in $P-X$. We then
let $\ell_{(v^{-}, v^{+})}:=0$ for every $v \in A_{i}$, set $X := X \cup A_{i}$,
increment $i$ by $1$, and repeat the above steps, until
$X = V$.  Computing the set $Y$, decrementing the flow, and finding the antichain $A_{i}$
are steps that can be done in time $O(m)$. Since we go through the main loop at most
$2n$ times, this implementation of the greedy algorithm runs in time $O(nm)$.

The  greedy point $\tilde x$ can be computed in time $O(n)$. The
corresponding potential $\tilde y$ can be found in $O(m)$ using a simple dynamic program.
The second execution of the greedy coloring algorithm can be done in time $O(n^2)$,
using the fact that the comparability graph of the interval order $I$ is a co-interval graph.
Finally, a bound on the complexity of the multiple selection procedure is $O(n^2)$.
So the whole algorithm runs in $O(nm) = O(n^3)$.

\section{Tightness}
\label{sec-discussion}

A natural question is whether there exists an algorithm for {\partsort} that
does at most $ITLB + O(n)$ comparisons between elements of $T$.
We show in this section that every algorithm that first extends the target poset to a weak order and then solves the problem on the weak order can be forced to make $ITLB + \Omega(n\log\log n)$ comparisons, both in the worst case and the average case. This is a consequence of the following theorem:

\begin{theorem}
\label{th-tight}
There exists a constant $c > 0$ such that, for all $n\geq 1$, there is
a poset $P$ on $n$ elements satisfying $H(W) \geq H(P) + c\,\log\log n$
for every weak order $W$ extending $P$.
\end{theorem}

\begin{figure}
\centering
\includegraphics[width=0.4\textwidth]{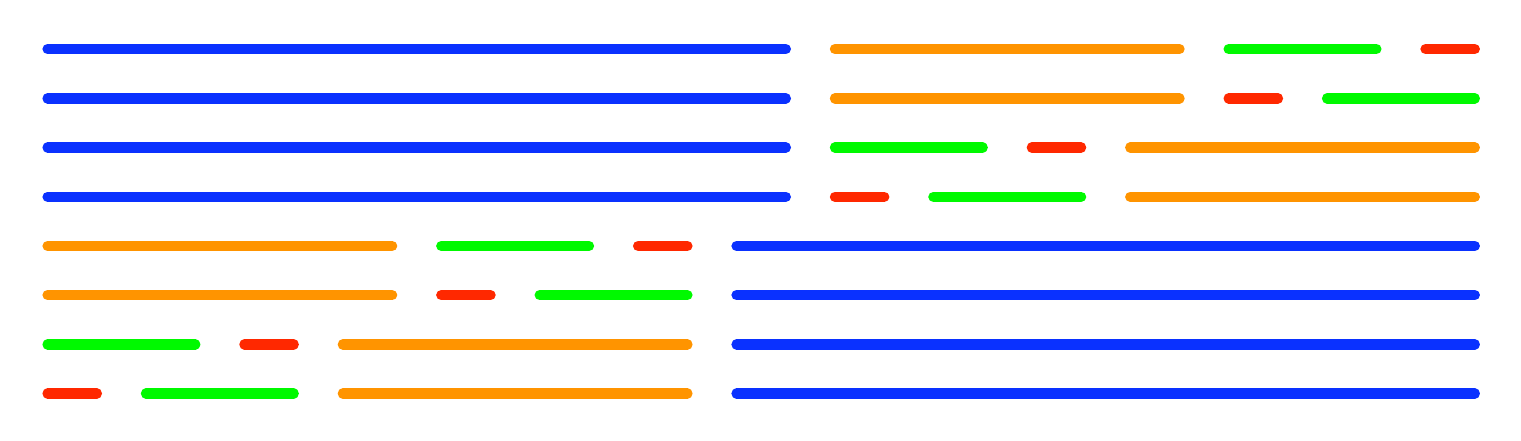}
\caption{\label{fig-tight} An interval representation of $G_{4}$ (colors highlight the different
levels of the construction).}
\end{figure}

In order to prove Theorem~\ref{th-tight}, we define a family $\{G_{k}\}$ ($k\geq 1$)
of interval graphs inductively as follows:
\begin{itemize}
\item $G_{1}$ consists of a unique vertex, and
\item for $k\geq 2$, the graph $G_{k}$ is obtained by first taking the disjoint union of
$K_{2^{k-1}}$ (the ``central clique'') with two copies of $G_{k-1}$, and then making half
of the vertices of the central clique adjacent to all vertices in the first copy, and the other half
to all those in the second copy.
\end{itemize}
It is easily seen that $G_{k}$ is indeed an interval graph, as is suggested
in Figure~\ref{fig-tight}. The graph $G_{k}$ has $k2^{k-1}$ vertices.
The complement $\bar G_{k}$ of $G_{k}$ is the comparability
graph of the interval order $I_{k}$ defined by an interval representation of $G_{k}$.

\begin{lemma}
\label{lem-tight-koerner}
$H(I_{k}) \leq (k+1)/2$.
\end{lemma}
\begin{proof}
By construction, the maximal stable sets of the graph $\bar G_{k}$ all have $2^{k-1}$ elements,
and there are $2^{k}-1$ such maximal stable sets.
We define a point $x^{(k)}$ of the stable set polytope $\STAB(\bar G_{k})$ as follows:
$$
x^{(k)} := \sum_{i=1}^{2^{k}-1}  \frac{1}{2^{k}-1}   \chi^{S_{i}},
$$
where $S_{1}, S_{2}, \dots, S_{2^{k}-1}$ are the maximal stable sets of $\bar G_{k}$. Observe that, for every $\ell \in \{0, \dots,  k-1\}$, there are $2^{k-1}$ vertices in $\bar G_{k}$ that belong to exactly $2^{\ell}$ different maximal stable sets (that is, there are $2^{k-1}$ intervals of each different length in the interval representation suggested in Figure~\ref{fig-tight}). We thus obtain the following upper bound on the entropy of $I_{k}$:
\begin{equation*}
H(I_{k})
\leq -\frac 1{k2^{k-1}}\sum_{\ell=0}^{k-1} 2^{k-1}\log{\frac{ 2^{\ell}}{2^{k}-1} }=\log{(2^k-1)}-\frac{k-1}{2}\leq (k+1)/2.
\end{equation*}
The lemma follows.
\end{proof}

We proceed by showing that every weak order extension of $I_{k}$ has relatively large entropy
compared to $I_{k}$. We first introduce some definitions.
Consider an arbitrary graph $G$ and a coloring $C_{1}, \dots, C_{\ell}$ of its vertices.
Similarly as how greedy points are defined (see Section~\ref{sec-perfect-graphs}),
one can associate an entropy to the latter coloring, namely, the entropy of the
probability distribution $\{ |C_i|\,/ n\}_{i = 1, \ldots, \ell}$:
$$
-\sum_{i=1}^{\ell} \frac{|C_{i}|}{n} \log \frac{|C_{i}|}{n}.
$$
The minimum entropy of a coloring is known as the {\em chromatic entropy} of $G$,
and is denoted by $H_{\chi}(G)$.
The chromatic entropy can be thought of as a constrained version of the
graph entropy, in which the stable sets involved in the definition of $H(G)$ are
required to form a partition of the vertices of $G$.

\begin{lemma}
\label{lem-chromatic-entropy}
Let $G$ be the comparability graph of a poset $P$.
Then any weak order extension $W$ of $P$ has entropy
$H(W)\geq H_{\chi} (G)$.
\end{lemma}
\begin{proof}
The maximal antichains of $W$ are pairwise disjoint, hence they correspond to a coloring of $G$.
The entropy of $W$ is equal to the entropy of the latter coloring, and thus is at least $H_{\chi}(G)$.
\end{proof}

Lemma~\ref{lem-chromatic-entropy} suggests finding a (good) lower bound on
$H_{\chi}(\bar G_{k})$, the chromatic entropy of $\bar G_{k}$. To achieve that, we
make use of the following result of~\cite{CFJ08-ALGO} (see Corollary~1 in that paper).

\begin{theorem}[\hspace{-.01em}\cite{CFJ08-ALGO}]
\label{th-ALGO}
Let $G$ be an arbitrary graph. Then the entropy of any coloring of $G$ produced by the
greedy coloring algorithm is at most $H_{\chi}(G) + \log e$.
\end{theorem}

We can therefore restrict ourselves to analyzing the entropy of greedy colorings of $\bar G_{k}$.
Recall that all maximal stable sets in $\bar G_{k}$ have the same cardinality $2^{k-1}$.
Consider the greedy coloring of $\bar G_{k}$ defined recursively as follows: take first
the stable set of cardinality $2^{k-1}$ that corresponds to the central clique in $G_{k}$, and
then, if $k\geq 2$, recurse on the two copies of $\bar G_{k-1}$ that are left.
Let $\tilde g_{k}$ denote the entropy of the resulting coloring of $\bar G_{k}$.

\begin{lemma}
\label{lem-tight-greedy}
$\tilde  g_{k} = (k-1)/2 + \log k$.
\end{lemma}
\begin{proof}
The greedy coloring defined above
consists of $2^{i-1}$ color classes of cardinality $2^{k-i}$, for $i=1,2,\dots, k$.
Hence, its entropy is
\begin{align*}
\tilde  g_{k} &= - \sum_{i=1}^{k} 2^{i-1} \cdot \frac{2^{k-i}}{k2^{k-1}} \log \frac{2^{k-i}}{k2^{k-1}}\\
&= \frac{1}{k} \sum_{i=1}^{k} \log \frac{k2^{k-1}}{2^{k-i}} \\
&= \frac{1}{k} \sum_{i=1}^{k} (\log k + (i-1))\\
&= \log k + \frac{k-1}{2},
\end{align*}
as claimed.
\end{proof}

We may now turn to the proof of Theorem~\ref{th-tight}.

\begin{proof}[Proof of Theorem~\ref{th-tight}]
Let $k\geq 1$ and consider the interval order $I_{k}$ defined above, of
order $n:=k2^{k-1}$.
Let also $W$ be an arbitrary weak order extending $I_{k}$.
Combining Lemmata~\ref{lem-tight-koerner},~\ref{lem-chromatic-entropy}
and~\ref{lem-tight-greedy} with Theorem~\ref{th-ALGO} gives
\begin{align*}
H(W) - H(I_{k})
&\geq H_{\chi}(\bar G_{k}) - H(I_{k})  \\
&\geq \left(\frac{k-1}{2} + \log k - \log e \right) - \frac{k+1}{2}\\
&= \log k - \log e - 1 \\\
&= \Omega(\log \log n),
\end{align*}
as claimed.
\end{proof}

\paragraph{Acknowledgments}
The authors wish to thank S\'ebastien Collette, Fran\c{c}ois Glineur and Stefan Langerman for useful discussions, and the anonymous referees for their comments on an earlier version of the paper.

\bibliographystyle{plain}

\end{document}